\documentclass[a4paper]{article}

\usepackage{INTERSPEECH2019}
\usepackage{graphicx}
\usepackage{amsmath}
\usepackage{amsthm}
\usepackage{caption}

\usepackage{enumitem}

\usepackage[labelformat=simple]{subfig}

\usepackage{hyperref}
\hypersetup{hidelinks}

\def\equationautorefname#1#2\null{
	Eq. (#2\null)
}

\title{Learn Spelling from Teachers: Transferring Knowledge from Language Models to Sequence-to-Sequence Speech Recognition}

\name{Ye Bai $^{1,2}$, Jiangyan Yi$^{1}$, Jianhua Tao$^{1,2,3}$,  Zhengkun Tian$^{1,2}$, Zhengqi Wen$^1$}
\address{$^1$NLPR, Institute of Automation, Chinese Academy of Sciences, China\\
	$^2$School of Artificial Intelligence, University of Chinese Academy of Sciences, China \\
	$^3$CAS Center for Excellence in Brain Science and Intelligence Technology}

\email{\{ye.bai, jiangyan.yi, jhtao, zhengkun.tian, zqwen\}@nlpr.ia.ac.cn}

\begin{document}

\maketitle
\begin{abstract}
	Integrating an external language model into a sequence-to-sequence speech recognition system is non-trivial. Previous works utilize linear interpolation or a fusion network to integrate external language models. However, these approaches introduce external components, and increase decoding computation. In this paper, we instead propose a knowledge distillation based training approach to integrating external language models into a sequence-to-sequence model. A recurrent neural network language model, which is trained on large scale external text, generates soft labels to guide the sequence-to-sequence model training. Thus, the language model plays the role of the teacher. This approach does not add any external component to the sequence-to-sequence model during testing. And this approach is flexible to be combined with shallow fusion technique together for decoding. The experiments are conducted on public Chinese datasets AISHELL-1 and CLMAD. Our approach achieves a character error rate of $9.3\%$, which is relatively reduced by $18.42\%$ compared with the vanilla sequence-to-sequence model.
\end{abstract}

\section{Introduction}

Attention based sequence-to-sequence (Seq2Seq) models have achieved promising performance in automatic speech recognition (ASR)  \cite{bahdanau2016end,chan2016listen,dong2018speech,chiu2018state}. A Seq2Seq model consists of two components: an encoder encodes the acoustic feature sequence into a high level representation, and a decoder generates the corresponding word sequence. The encoder leverages attention mechanism to fuse extracted features into a fixed-dimensional vector for capturing global semantic information of a speech signal. The decoder is a conditional language model (LM) to capture linguistic information of transcriptions. During decoding stage, the decoder predicts the current word in terms of the acoustic encoding of the encoder, history context, and the previous word at each step. This architecture is also referred to as Listen, Attend, and Spell \cite{chan2016listen}.

Compared with speech transcriptions, abundant unsupervised text corpora, which have rich linguistic information, are easier to obtain. Large scale external text data is commonly used to train language models (LMs) to improve ASR performance in conventional hidden Markov model (HMM) or connectionist temporal classification (CTC) based ASR pipelines. However, because the encoder and the decoder are optimized jointly, it is non-trivial to integrate an external LM into a Seq2Seq model.

Shallow fusion and deep fusion are two approaches to integrating an LM into a Seq2Seq model \cite{gulcehre2015on}. Shallow fusion performs log-linear interpolation between the decoder of a Seq2Seq model and an external LM during beam search. The external LM can be $n$-gram LM or neural network language models (NNLMs). It has achieved success in ASR tasks \cite{bahdanau2016end,kannan2018an}. Various deep fusion approaches leverage a neural network to fuse hidden representations of the Seq2Seq decoder and the external neural network based LM \cite{gulcehre2015on}. Cold fusion and component fusion utilize a pre-trained recurrent neural network language model (RNNLM) and gating mechanism to improve ASR performances \cite{sriram2018cold,shan2019component}. These fusion approaches have shown promising performance. However, the neural network of the external LM increases complexity of the Seq2Seq model. Specifically, the fusion network introduces external parameters into the Seq2Seq model for deep fusion. Both shallow fusion and deep fusion need the external LM during test stage. It introduces external complexity into the ASR system.

We propose a knowledge distillation (KD) \cite{hinton2015distilling} based training approach to integrating an external LM into a Seq2Seq model. First, an RNNLM is trained on large scale text data. Then, the RNNLM is used to generate soft labels of speech transcriptions to train the Seq2Seq model. This training approach is also named as Teacher/Student model: the teacher (RNNLM) provides soft labels as prior knowledge to ``teach'' the student (Seq2Seq decoder). Thus, we refer to the proposed training approach as ``Learn Spelling from Teachers'' (LST). LST is simple to implemented: it does not modify the model structure, and only needs to train an RNNLM to generate soft labels. With LST, the external LM is only needed during training, so it does not increase complexity of the model for testing. Furthermore, LST and shallow fusion can be used together to achieve better performance. We conduct experiments on publicly available AISHELL-1\footnote{http://openslr.org/33/} dataset \cite{bu2017aishell} and CLMAD\footnote{http://openslr.org/55/} text dataset \cite{yebai2018clmad} to show the effectiveness of the proposed LST. We use Speech-Transformer \cite{dong2018speech} as the backbone network. Our proposed approach reduced character error rate (CER) from $11.4\%$ to $9.3\%$. we further utilized shallow fusion for the model trained with LST, and achieved CER of $8.3\%$. 
 
The rest of this paper is organized as follows. \autoref{sec:bg} introduces the background. \autoref{sec:methods} introduces the proposed LST. \autoref{sec:rw} introduces the related work. \autoref{sec:exp} describes the experimental results. \autoref{sec:conc} summarizes the paper.

\begin{figure*}[!t]\centering
	\subfloat[Vanilla Seq2Seq Model] 
	{ \label{fig:enc_dec}
		\includegraphics[width=0.5\columnwidth]{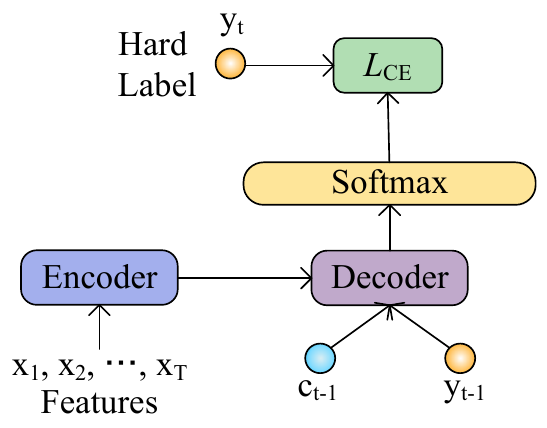}
	}		
    \qquad
	\subfloat[Learn Spelling from Teachers (LST)]  
	{ \label{fig:kd}
		\includegraphics[width=1.0\columnwidth]{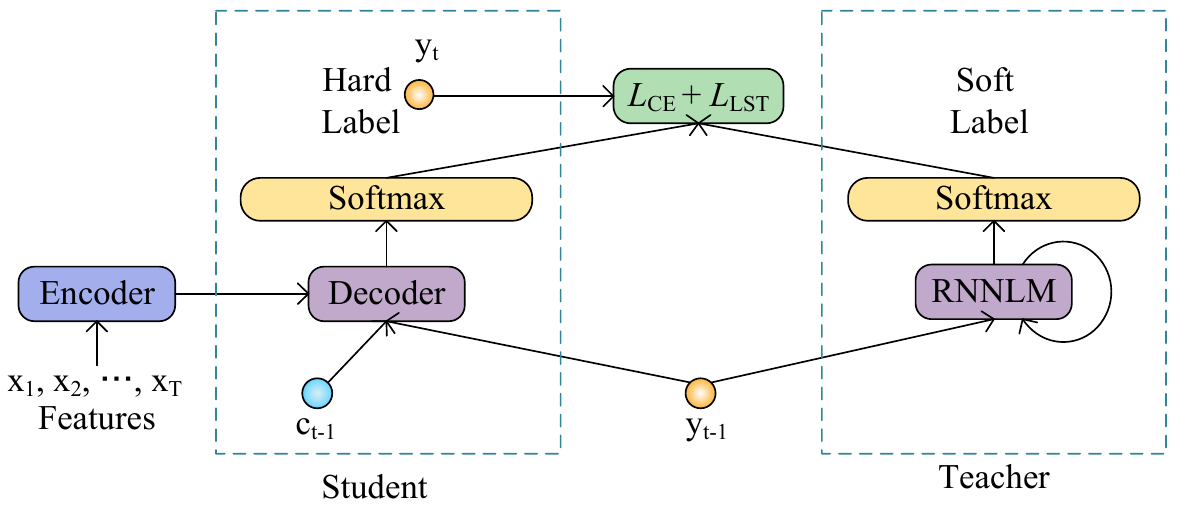}
	} 
	\caption{(a) illustrates a basic encoder-decoder architecture for ASR. $\text{x}_1, \cdots, \text{x}_t $ represent acoustic features, $\text{c}_{t-1}$ denotes the context of $t-1$ step, and $\text{y}_{t-1}$ denotes the previous token. The decoder predicts the current token in terms of the context $\text{c}_{t-1}$, the previous token $\text{y}_{t-1}$, and the acoustic vector generated by the encoder. The loss is computed with the softmax function of the decoder and the current ground truth token $\text{y}_{t}$. (b) illustrates the proposed ``Learn Spelling from Teachers'' (LST) approach. The RNNLM generates soft labels to train the Seq2Seq model, and it is removed during testing.} 
	\vspace{-15pt}
\end{figure*}

\vspace{-5pt}
\section{Background: Seq2Seq models for ASR}
\label{sec:bg} 

A basic Seq2Seq model is shown in \autoref{fig:enc_dec}. First, a speech signal is processed into an acoustic feature sequence. Then, an encoder network encodes the sequence into a high level acoustic representation. The encoder can be a recurrent neural network \cite{chan2016listen,chorowski2015attention} or a transformer \cite{dong2018speech}. The decoder is a conditional LM: given the high level acoustic representation, the previous token, and history context, it predicts the current token. The probability distribution on the vocabulary is computed by a softmax function. 

The attention is an important mechanism to capture the relationship between the acoustic representations and the current state of the decoder. The attention scores are computed in terms of the current state of the decoder and the high level acoustic representations, and then the acoustic information and the decoder state are fused.

The encoder and decoder are trained jointly. The training criterion is cross entropy:
\begin{equation}
\label{eq:loss1}
L_{\text{CE}}(\theta) = -\sum_{k=1}^{K} \delta(k, y_{t}) \log P_{\text{S2S}}(k | y_{t-1}, c_{t-1}, \bm{x}; \theta), 
\end{equation}
where $k$ is the index of each token, $K$ is the vocabulary size, $y_t$ is the index of the corresponding ground truth token at step $t$, $y_{t-1}$ is the previous token, $c_{t-1}$ is the history context, $\bm{x}$ is the acoustic features, $P_{\text{S2S}}$ represents probability, and $\theta$ stands the parameters of the whole network. $\delta(\cdot, \cdot)$ is $1$ if the two variables are equal, and $0$ otherwise.

\vspace{-5pt}
\section{Distilling knowledge from external LMs}
\label{sec:methods}

\begin{figure}[!t]\centering
	\includegraphics[width=0.75\columnwidth]{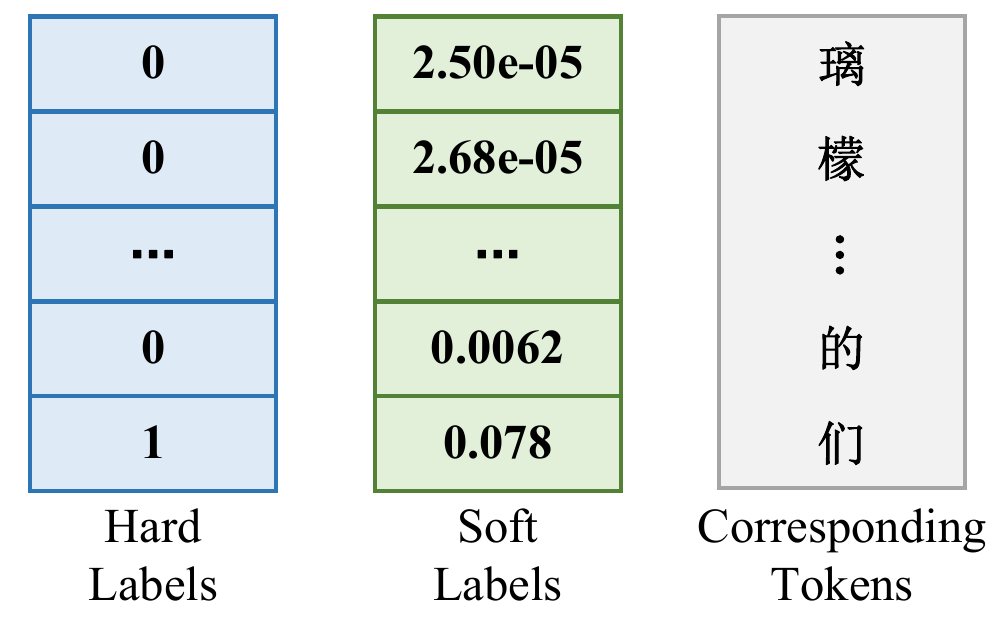}
	\caption{Hard labels and soft labels at one time step of a sequence for training. The values of the soft labels reflect knowledge of the external LM.} 
	\label{fig:sl}
	\vspace{-15pt}
\end{figure}

The basic idea of ``Learn Spelling from Teachers'' (LST) is: first, train an RNNLM on an external large scale text corpus, and then use this RNNLM to guide Seq2Seq model training. Besides $1$-of-K hard labels provided by the transcriptions, the RNNLM provides soft labels, which carries the knowledge of the text corpus. The soft labels are probabilities estimated by the RNNLM. \autoref{fig:sl} shows the hard labels and soft labels of tokens in the vocabulary at one time step in a sequence. The soft labels contain more information than hard labels, e.g. some tokens have relatively large probabilities, and some tokens have very small probabilities.

Given the context and the previous token, the probability of $k$-th token in vocabulary estimated by the RNNLM is
\begin{equation}
	\label{eq:softmax_lm}
	P_{\text{LM}}(k | y_{t-1}, h_{t-1}) = \frac{\text{exp}(z_k / T)}{\sum_{i=1}^{K}\text{exp}(z_i / T)},
\end{equation}
where $z_i$ is $i$-th node of latent variable before the softmax function, $K$ is the vocabulary size, $y_{t-1}$ is the previous token, $h_{t-1}$ is the history context, and $T$ is a parameter called temperature to smooth the outputs.

To make the Seq2Seq model learn the knowledge from the RNNLM, we minimize the Kullback-Leibler divergence (KLD) between estimated probability of the RNNLM $P_{\text{LM}}$ and the estimated probability of the Seq2Seq model $P_{\text{S2S}}$. Let $P_{\text{S2S}}^k = P_{\text{S2S}}(k|y_{t-1}, c_{t-1}, \bm{x}; \theta)$, and $P_{\text{LM}}^k = P_{\text{LM}}(k | y_{t-1}, h_{t-1})$, the KLD is
\begin{equation}
D_\text{KL}(P_\text{LM}||P_\text{S2S}) =-\sum_{k=1}^K P_{\text{LM}}^k \log \frac{P_{\text{S2S}}^k}{P_{\text{LM}}^k}.
\end{equation}
Because $P_{\text{LM}}$ is fixed during training the Seq2Seq model, the loss function is equivalent to the cross entropy form:
\begin{equation}
\label{eq:loss2}
L_{\text{LST}}(\theta) = -\sum_{i=1}^{K} P_{\text{LM}}^k \log P_{\text{S2S}}^k.
\end{equation}
We refer to the above loss as LST loss.

The cross entropy loss in \autoref{eq:loss1} and the LST loss in \autoref{eq:loss2} are weighted with a coefficient $\lambda \in [0,1] $, then the final loss is  
\begin{equation}
\label{eq:loss}
L(\theta) =\lambda L_{\text{CE}}(\theta) + (1-\lambda) L_{\text{LST}}(\theta).
\end{equation}

We can simplify the above equation as the label interpolation form:
\begin{equation}
\label{eq:loss_form2}
L(\theta) = -\sum_{k=1}^{K}(\lambda \delta(k, y_{t}) + (1-\lambda)P_{\text{LM}}^k ) \log P_{\text{S2S}}^k.
\end{equation}
Thus, compared with the vanilla Seq2Seq model, we just modify the labels rather than the loss function during training stage. $L(\theta)$ combines the knowledge from transcriptions and the knowledge from the LM. The LST is illustrated in \autoref{fig:kd}. 

Comparing \autoref{fig:kd} with \autoref{fig:enc_dec}, we can see that LST is only used for training, and the external RNNLM is removed during testing. So the computation is the same as the original Seq2Seq for testing. In order to achieve better performance, shallow fusion can be further used with LST during decoding. In addition, besides ASR, our proposed LST can be generally used for Seq2Seq models. 

\vspace{-5pt}
\section{Related work}
\label{sec:rw}

\textbf{Knowledge distillation}. KD was proposed for model compression \cite{hinton2015distilling}. It is also referred to as teacher-student learning. Yoon et al. proposed to use KD to reduce the size of a Seq2Seq model for machine translation \cite{kim2016sequence}. It has also been used for domain adaptation for acoustic models \cite{li2017large} and language models \cite{andres2018efficient}. Different from these work, our work focuses on integrating external language models for Seq2Seq ASR systems.\\ 
\textbf{Label smoothing}. Label smoothing have been used to prevent the Seq2Seq ASR model making overconfident predictions \cite{chiu2018state,dong2018speech,chorowski2017towards}. It can be seen as a special case of KLD regularization when assuming the prior label distribution is uniform \cite{pereyra2017regularizing}. Unlike label smoothing, LST leverages an RNNLM to provide a context-dependent prior distribution rather than a simple uniform distribution. Instead of assumption, the prior distribution is estimated with a data-driven method. Besides solving the overconfident problem, LST introduces knowledge from an external large scale text data corpus.

\vspace{-5pt}
\section{Experiments}
\label{sec:exp}
\vspace{-5pt}
\subsection{Datasets}
\vspace{-5pt}

We use a Chinese corpus AISHELL-1 to evaluate our proposed approach \cite{bu2017aishell}. The training set contains $150$ hours of speech ($120,098$ utterances) recorded by $340$ speakers. The development set contains $20$ hours of speech ($14,326$ utterances) recorded by $40$ speakers. And the test set contains $10$ hours of speech ($7,176$ utterances) recorded by $20$ speakers. The speakers of the training set, development set, and test set are not overlapped. All the recordings of the corpus are in $16$ kHz WAV format. The content of the speech is news with different topics. 

A subset of CLMAD \cite{yebai2018clmad,li2007scalable} text dataset is used as external text dataset\footnote{This subset of the external text has been shared with OneDrive: https://1drv.ms/u/s!An08U7hvUohBb234-V-Z0Qb\_Zcc}. We use an open source tool XenC to extract the subset of CLMAD which is topic matched with AISHELL-1 \cite{rousseau2013xenc}. The preprocessing steps are as follows:
\begin{enumerate}[label=(\arabic*)]
	\item Select $3$ million sentences which have small cross entropy with AISHELL-1 training transcriptions \cite{moore2010intelligent}; 
	\item Remove the sentences whose lengths are longer than $50$;
	\item Mix the remained sentences with training transcriptions (which are duplicated $10$ times to improve proportion);
	\item Re-segment the word sequences into characters.
\end{enumerate}
The information of the text data is shown in \autoref{tab:text_data}.

\vspace{-5pt}
\subsection{Experimental setup}
\vspace{-5pt}
In this paper, we employ Speech-Transformer \cite{dong2018speech,zhou2018syllable}, a non-recurrent Seq2Seq model for speech recognition, as the backbone network. Instead of hidden states and recurrent structures of RNNs, the transformer models the context by computing attention directly. Please see \cite{dong2018speech,chorowski2015attention,zhou2018syllable} for details of the transformer.


The acoustic features are $80$-dimension Mel-filter bank features (FBANK), which are extracted every 10ms with 25ms of frame length. Each frame is spliced with three left frames. So, the input of the network is $320$-dimensional. The sequence is subsampled every three frames. The Speech-Transformer consists of $6$ blocks of an encoder and $6$ blocks of a decoder. The dimensionality of the model is $512$, and the number of inner nodes of the fully connected feed-forward network is $2048$. The number of heads is $8$. The modeling units of the decoder are $4232$ characters, including three special symbols ``$<$unk$>$'', ``$<$sos$>$'', ``$<$eos$>$'', which represent unknown character, start of a sentence, end of a sentence, respectively. The character embeddings is shared with the output weights of the decoder \cite{press2016using}. Following \cite{dong2018speech}, we use Adam optimizer with $\beta_1 = 0.9$, $\beta_2 = 0.98$, $\epsilon = 10^{-9}$. The learning rate $\alpha$  is updated as follows:
\begin{equation}
	\alpha = k \cdot d_{\text{model}}^{-0.5}  \cdot \text{min} (n^{-0.5}, n \cdot warmup^{-1.5}), 
\end{equation}
where $d_{model}$ is the dimensionality of the model, $n$ is the step number, $k$ is a tunable parameter, and the learning rate increases linearly for $warmup$ steps. We set $k=0.5$, $warmup=8000$. The model is trained for $50$ epochs. There are utterances containing about $20$K frames in one batch. The development set is used for validation. Only the model which achieves the lowest cross entropy on development set is stored as the final model.

The external RNNLM is a two layers of long short-term memory (LSTM) network. The modeling units are the same as the Seq2Seq model. The RNNLM is trained on the external text. The embedding size of the RNNLM is $300$, and the number of LSTM cells of each layer is $1024$. The RNNLM is trained on external text. The stochastic gradient decent (SGD) with momentum as the optimizer for training the RNNLM. The momentum is set to $0.9$, and the learning rate is set to $1.0$. The RNNLM is trained for $5$ epochs. 

For decoding, we set beam width to $5$ for beam search, and maximum decoding length to $60$.


\begin{table}[!t]	
	\caption{The description of the text.}
	\centering
	\begin{tabular}{ c  c  c  c } 
		\toprule
		& \#Sentences     &    \#Characters    &    Size              \\
		\midrule
		Training Trans.  & $120,098$    & $1,730,113$   &    $6.6$MB           \\
		Test Trans. & $7,176$    & $104,765$   &    $0.4$MB           \\
		External Text  &  $3,703,982$    &  $75,893,998$ &   $290$MB            \\
		\bottomrule
	\end{tabular}
	\label{tab:text_data}	
	\vspace{-15pt}	
\end{table}

\vspace{-5pt}
\subsection{Results and analysis}
\vspace{-2.5pt}
\subsubsection{The effectiveness of external text}
\vspace{-2.5pt}
	
Firstly, we demonstrate the effectiveness of the external text data and the RNNLM. We compute the perplexities on AISHELL-1 test transcriptions, which are shown in \autoref{tab:perplexity}. Note that the data is in character level, so the perplexities are relatively smaller. We can see that compared with the $3$-gram with Kneser-Ney smoothing trained on training transcriptions, the $3$-gram trained on external text achieves a significant reduction of perplexity. Moreover, the RNNLM achieves about a $28\%$ relative reduction over the $3$-gram trained on the external text.

\vspace{-5pt}
\subsubsection{The impact of hyper-parameters}
\vspace{-2.5pt}

\autoref{tab:ab1} shows the the character error rates (CERs) on development the set with different temperature $T$ in \autoref{eq:softmax_lm} when $\lambda$ is fixed at $0.9$. The temperature controls the smoothness of the soft labels generated by the RNNLM. When it is too small, the soft labels are too sharp, and the Seq2Seq training is perturbed heavily. When it is too large, the soft labels are too smoothed to affect the training. We can see that when the temperature is set to $5.0$, the model achieves the best performance. 

Then we fix $T$ at $5.0$ and evaluate the influence of $\lambda$. The parameter $\lambda$ controls the proportion of the ground truth hard labels to the soft labels of the RNNLM. The results are shown in \autoref{tab:ab2}. We can see that when $\lambda=0.9$, the model achieves the best performance on the development set. According to the above results in \autoref{tab:ab}, we select $\lambda=0.9$ and $T=5.0$ as the final hyper-parameters. We refer to the model trained with $\lambda=0.9$ and $T=5.0$ as ``Seq2Seq+LST'' in the rest.


\begin{table}[!t]
	\caption{The perplexities on transcriptions of AISHELL-1 development set.}
	\vspace{-5pt}	
	\centering
	\begin{tabular}{ l  c  } 
		\toprule
		\multicolumn{1}{c}{LM}          &    PPL  \\
		\midrule
		$3$-gram (Training Trans.)  &   $70$  \\
		$3$-gram (Ext. Text) &     $47$ \\
		RNNLM (Ext. Text)    &     $34$  \\
		\bottomrule
	\end{tabular}
	\label{tab:perplexity}	
	\vspace{-5pt}
\end{table}


\begin{table}[!t]
	\caption{Comparisons of different hyper-parameters on the development set.  }
	\vspace{-7.5pt}
	\centering
	\subfloat[Varing $T$ for RNNLM softmax ($\lambda=0.9$).]{
		\centering		
		\begin{tabular}{  l  c  } 
			\toprule
			Temperature &    CER\%  \\
			\midrule
			$T=1.0$  & $10.5$   \\
			$T=3.0$  & $8.3$   \\
			$T=5.0$  & $8.0$   \\
			$T=7.0$  & $8.2$   \\
			$T=10.0$ & $8.2$  \\
			$T=15.0$ & $9.0$ \\
			\bottomrule
		\end{tabular}	
		\label{tab:ab1}	
	}
	\qquad\quad
	\vspace{-7.5pt}
	\subfloat[Varing $\lambda$ for interpolation ($T=5.0$).]{
		\centering	
		\begin{tabular}{  l  c  } 
			\toprule
			Weight&    CER\%  \\
			\midrule
			$\lambda=0.3$  &   $9.4$   \\
			$\lambda=0.5$  &   $9.2$  \\
			$\lambda=0.8$  &   $9.2$  \\
			$\lambda=0.85$ &     $8.9$ \\
			$\lambda=0.9$  &     $8.0$  \\
			$\lambda=0.95$ &     $9.3$  \\
			\bottomrule
		\end{tabular}
		\label{tab:ab2}	
	}
	\label{tab:ab}
	\vspace{-20pt}	
\end{table}

\vspace{-5pt}
\subsubsection{The effectiveness of the proposed approach}
\vspace{-2.5pt}
\autoref{tab:comparison} gives the results on the test set of each model. ``Seq2Seq'' is the plain Seq2Seq without regularization. Compared to ``Seq2Seq'', ``Seq2Seq+LST'' achieves an $18.42\%$ relative reduction in character error rate. 

We report results of two KLD based regularization approaches, namely label smoothing and unigram smoothing. For label smoothing, the prior label distribution is assumed to be a uniform distribution. The label smoothing achieves a $4.4\%$ relative reduction in character error rate. For unigram smoothing, the prior label distribution is assumed to be the frequency of each label. The frequency is estimated on the external text. Because the unigram is too sharp, it introduces noises and affects training. We add $0.1$ to the frequencies and re-normalize them to smooth the unigram. We can see that the original unigram hurts the performance, and the smoothed unigram improves the performance. Both label smoothing and unigram smoothing are effective for regularizing the model. The unigram should be smoothed for training to reduce the sharp problem.

From \autoref{tab:comparison}, we can see that ``Seq2Seq+LST'' outperforms both label smoothing and unigram smoothing (without shallow fusion). We analyze that the assumptions of label smoothing (uniform distribution) and unigram smoothing (unigram frequency) do not match real situations. However, LST, which is a data-driven approach, does not assume prior distributions.

We further leverage shallow fusion with the RNNLM for each model. The weight of LM is $0.1$. The RNNLM is the same one which is used for LST. We can see that shallow fusion improves performances for all models. ``Seq2Seq + LST'' model outperforms ``Seq2Seq + Label Smoothing + SF'' model (CER $9.9\%$), which demonstrates that LST is an effective way to improve the performance of Seq2Seq models. Moreover, the model which uses LST and shallow fusion together, i.e. ``Seq2Seq + LST + SF'', achieves the best CER of $8.3\%$.


\begin{table}[!t]
	\caption{Comparisons on the test set. LST represents our proposed ``Learn Spelling from Teachers'' approach. SF means using shallow fusion during decoding. }
	\centering
	\vspace{-5pt}	
	\begin{tabular}{ l  c  } 
		\toprule
		\multicolumn{1}{c}{Model} &    CER\%  \\
		\midrule
		Seq2Seq  (baseline)            &   $11.4$    \\ 
		Seq2Seq + SF            &   $10.5$    \\ \cmidrule(lr){1-2}
		Seq2Seq + Label Smoothing    &   $10.9$    \\
		Seq2Seq + Label Smoothing + SF & $9.9$       \\ \cmidrule(lr){1-2}
		Seq2Seq + Original Unigram Smoothing & $15.7$   \\
		Seq2Seq + Smoothed Unigram Smoothing  & $10.0$   \\
		Seq2Seq + Smoothed Unigram Smoothing  +  SF & $8.7$    \\ \cmidrule(lr){1-2}	 
		Seq2Seq + Proposed LST                         &   $9.3$     \\
		Seq2Seq + Proposed LST + SF   &   $8.3$     \\
		\bottomrule
	\end{tabular}
	\label{tab:comparison}	
	\vspace{-10pt}	
\end{table}

\begin{figure}[!t]%
	\centering
	\includegraphics[width=1.0\linewidth]{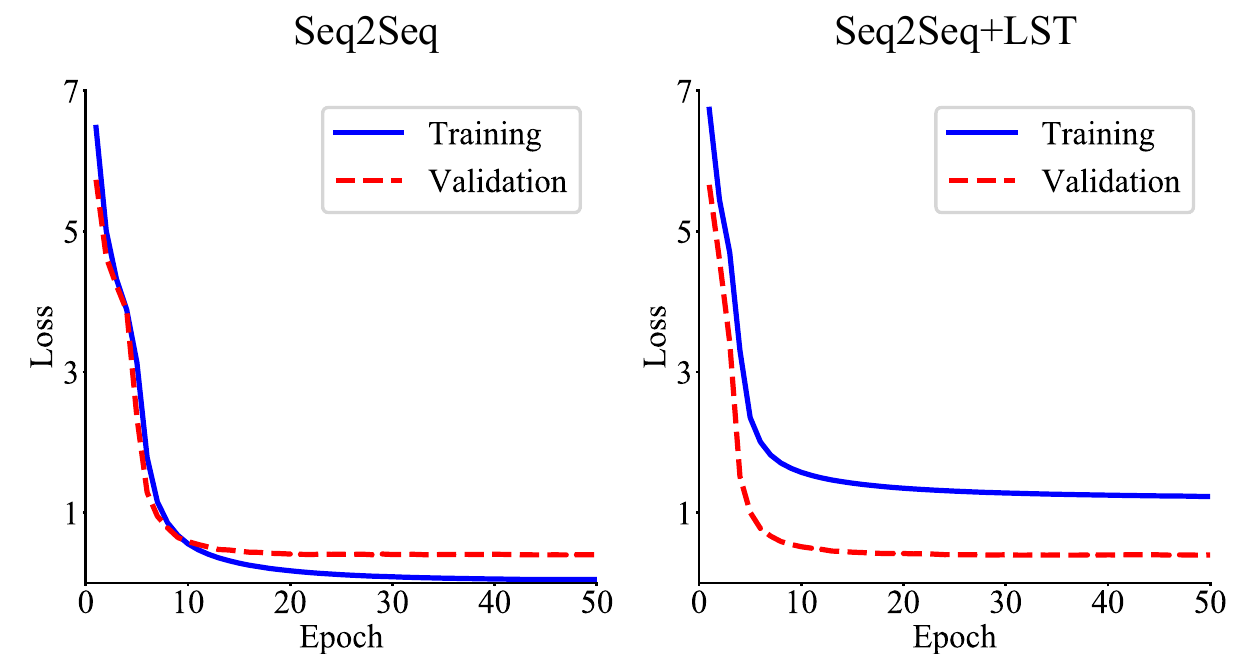}
	\vspace{-20pt}
	\caption{The loss curves of Seq2Seq model (left) and Seq2Seq model with LST (right). For Seq2Seq model, the training loss is lower than validation loss. However, with LST, the training loss is higher than the validation loss. Moreover, the validation loss in the right figure is a little smaller than the left one.}
	\label{fig:loss}
	\vspace{-20pt}
\end{figure}


To further show the effect of our proposed approach, we draw the loss curves with baseline ``Seq2Seq'' and the proposed ``Seq2Seq+LST'' in \autoref{fig:loss}. For ``Seq2Seq'' model, the training loss is lower than validation loss. However, for ``Seq2Seq+LST'', the training loss is higher than validation loss. The final validation loss of ``Seq2Seq+LST'' is a little bit smaller than ``Seq2Seq''. This result shows regularization effect of LST.

\vspace{-5pt}
\section{Conclusions}
\vspace{-5pt}

\label{sec:conc}

In this paper, we propose LST training approach to integrating an external RNNLM into a Seq2Seq model. An RNNLM is first trained on large scale external text data. Then, the RNNLM provides soft labels of training transcriptions to train the Seq2Seq model. We used transformer based Seq2Seq as backbone, and conducted experiments on public available Chinese datasets AISHELL-1 (speech) and CLMAD (external text). The experiments demonstrate the effectiveness of our proposed approach. We will try integrating more powerful language models into Seq2Seq systems in the future.

\vspace{-5pt}
\section{Acknowledgements}
\vspace{-5pt}
This work is supported by the National Key Research \& Development Plan of China (No.2017YFB1002801), the National Natural Science Foundation of China (NSFC) (No.61425017, No.61831022, No.61773379, No.61603390), the Strategic Priority Research Program of Chinese Academy of Sciences (No.XDC02050100), and Inria-CAS Joint Research Project (No.173211KYSB20170061).

\bibliographystyle{IEEEtran}

\bibliography{kd_extlm}


\end{document}